# Heterogeneous Runtime Verification of Safety Critical Cyber Physical Systems

Smitha Gautham, Abhilash Rajagopala, Athira Varma Jayakumar, Christopher Deloglos,
Erwin Karincic and Carl Elks

Virginia Commonwealth University

*Abstract*— Advanced embedded system technology is one of the key driving forces behind the rapid growth of Cyber-Physical System (CPS) applications. Cyber-Physical Systems are comprised of multiple coordinating and cooperating components, which are often software intensive and interacting with each other to achieve unprecedented tasks. Such complex CPSs have multiple attack surfaces and attack vectors that we have to secure against. Towards this goal, we demonstrate a multilevel runtime safety and security monitor framework where there are monitors across the CPS for detection and isolation of attacks. We implement the runtime monitors on FPGA using a stream-based runtime verification tool called TeSSLa. We demonstrate our monitoring scheme for an Autonomous Emergency Braking (AEB) CPS system.

## I. INTRODUCTION

Cyber Physical Systems (CPS) are used in a number of safety critical applications and their complexity and functionality are increasing rapidly. Applications such as robotics, autonomous systems, telemedicine and avionics are examples of CPSs becoming more complex, incorporating new technology and with varied features. For example, a CPS such as a modern automobile has environmental and engine control systems, sensors of all types, navigation systems, cameras all working together to achieve enhanced driving experiences and safety [1]. With the emergence of autonomous vehicle operations, the sphere of control with respect to vehicle is increased to include awareness and sensing of other traffic. As such, the functional and structural complexity of these systems is rapidly increasing which can have significant engineering and operational impacts.

With the evolution of technology and emergence of autonomous systems and Internet of Things (IOT), new challenges such as attacks, faults can occur at multiple levels. This necessitates a means of ensuring that the system requirements and design assurance are carried further down to the implementation stages. A potential solution is runtime monitors that observe system behavior and provide assurance of safety and security without interfering with the system under observation [1].

Modern embedded digital devices have evolved to the point where all types of heterogeneous processing and data mobility reside within a single platform or chip from low-level onboard sensor pre-processors to dedicated network communication cores. The integration of customizable system on a chip technology and flexible communication enables tight integration with the physical world. However, CPS architected from this technology are increasingly vulnerable to design flaws, software flaws, and security threats at multiple levels that span both hardware and software implementations. Given that computation and data processing vulnerabilities may exist at multiple levels in embedded CPS, it follows that solutions should be present at the levels where the faults or vulnerabilities originate. We assert that a viable approach to this problem is to employ runtime security and safety monitoring at these various levels of processing and integration.

Our heterogeneous multilevel runtime monitors placed across the CPS, observe streams of information coming from the CPS and verify that the operation is within the expected safety and security bounds. We implement our runtime monitors using a stream-based verification tool called TeSSLa [2]. The runtime monitors are on a separate platform and isolated from the system being monitored to ensure that they do not interfere with the CPS. Our target CPS under observation is an Autonomous Emergency Braking (AEB) controller, which is a part of the Advanced Driver Assistance Systems (ADAS) found in most modern vehicles [3]. The functioning of the autonomous features can be affected by attacks on the sensors, attack on the networks through which the AEB communicates with other Electronic Control Units (ECU) in a vehicle or attack on the AEB computational unit itself. In order to detect and isolate attacks in CPSs we propose a multilevel monitor framework.

The contributions of this paper are:

*1)* A novel multilevel monitoring framework for runtime safety and security monitoring of CPSs.
*2)* Implementation of runtime monitors on FPGA using stream based runtime verification tool

## II. RELATED WORK

A CPS is often comprised of numerous integrated components and subsystems interacting and communicating with each other to satisfy system level goals. These goals are often related to the functional performance, safety and security of the service a CPS is providing – example being, an automobile cruise control that will always disengage when the brake is applied. Numerous work in literature have considered runtime monitors to ensure such operational safety and security properties of CPS. Reference [4] presents a survey of runtime monitoring architectures in distributed real time systems by providing three monitor architectures namely BUS, single process and distributed process architectures. Reference [5] provide a runtime monitoring framework to monitor black-box components in COTS processors. They use the bus monitor architecture where an external monitor can be used as a system component, which can silently receive messages over a system BUS without intrusion. Reference [6] integrates run time monitoring in an automotive development workflow and explores using this in autonomous systems. Reference [7] uses safety guards that are inbuilt runtime enforcers that ensure that the system satisfies predefined properties even under malicious attack. These papers emphasize the need for runtime monitors for safety critical CPSs.

Having multiple local and global monitors in a vehicle to ensure safety is proposed by [8]. A comprehensive review of various attacks surfaces, safety failures present in hardware, communication and processing levels for autonomous vehicles is presented in [9] and [10]. This motivates our proposed idea of having multiple monitors across various levels in a CPS to detect attacks.

### III. FORMAL DEVELOPMENT OF MONITORING FRAMEWORK

We discuss typical attacks on a CPS to motivate the development of the monitoring framework. We then describe the formalism for a single monitor, extend it to multilevel monitors and finally explain the necessity for such a multilevel monitoring scheme. We also present a threat model for Autonomous Emergency Braking (AEB) System, which is the representative system for monitoring in our paper.

*A. Attacks on CPS*

Typically, the attacks on a CPS can often be classified into three domains: **a.** Low level attacks on *hardware/firmware* devices. These include attacks on sensor (e.g. sensor spoofing that could result in signal delay or missed information, firmware attacks etc.) or actuator attacks. **b.** Attacks on *connection/network* layer (e.g. CAN Bus in cars) that include Denial of Service (DoS), packet injection, eavesdropping, etc. **c.** Attacks on the *computational elements* (e.g. malware injection, control flow attack, buffer overflow etc.) that affect the proper functionality of any CPS subsystem.

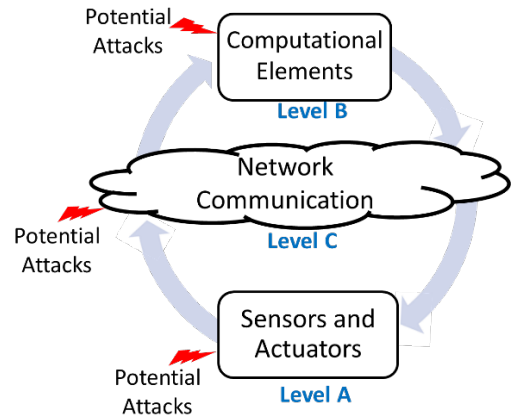

Fig. 1: Attack surface in Cyber Physical Systems.

Given such attacks can happen at multiple levels in a CPS as shown in Fig. 1, it is necessary to have monitors at multiple levels to ensure more classes of faults/attacks can be detected and isolated before they propagate through the system causing disastrous consequences.

*B. Single Monitor Model*

A monitor $M$ observes streams of information from the CPS. Such information could be discrete or continuous time states, sensor data or other states that we view from a historical perspective i.e. starting from the current state of the CPS to states previous in time, or past temporal observations. Each stream can be a sequence of inputs or outputs to and from the CPS or sub-system within the CPS. We denote the $k^{th}$ prefix for past $m$ instances of an infinite stream, $s$ as

$s^k = (s(k-m), \ldots, s(k-2), s(k-1), s(k))$

Hence, at time $k$ the stream contains information of the past $m$ instances starting with $s(k-m)$ and ending at the current instance $(k)$. The language of the monitor is defined by the set of monitored streams,

$w = ins^k, os^k, S_1^k, S_2^k, \ldots, S_m^k$, where

- $ins^k$ the input stream,
- $os^k$ the output stream,
- $S_1^k, S_2^k, \ldots, S_m^k$ the states of the system from current time $k$ for the past $m$ instances. We note that it sometimes suffices to monitor a specific relationship between the input and output stream to ensure there are no faults or security violations. In such cases, we may not need to deduce any specific internal state of the system.

Assuming that all security violations or faults in the system are observable in the monitored stream $w$ (which could be an input/output stream or state), the monitor $M$ makes safety and security assessments on the CPS based on a detection predicate. We extend the single monitor framework to multiple monitors to ensure timely detection of attacks at the point of vulnerability.

## C. Monitors at Multiple Levels

The single monitor $M$, is extended to comprise of monitors at multiple levels (Fig. 2) and denoted as the set $M = (M1, M2, …, Mn)$. Each monitor looks at different streams of information coming from the CPS or its subsystem. Thus, the monitored stream will be
$w_1 = ins^k{}_1, os^k{}_1, S_1{}^k{}_1, S_2{}^k{}_1, …, S_m{}^k{}_1$
$w_2 = ins^k{}_2, os^k{}_2, S_1{}^k{}_2, S_2{}^k{}_2, …, S_m{}^k{}_2 ….$
$w_n = ins^k{}_n, os^k{}_n, S_1{}^k{}_n, S_2{}^k{}_n, …, S_m{}^k{}_n$
Where $w_1, w_2, … w_n$ are the streams monitored by 1st, 2nd, …nth monitor.

In our simplified monitor architecture in this paper, we consider monitors at two levels, $M1$ (Data Monitor in Fig. 2) verifying the information integrity of the CPS, and $M2$ (Functional Monitor in Fig. 2) verifying the overall functionality of a sub-system of the CPS such as the control unit.

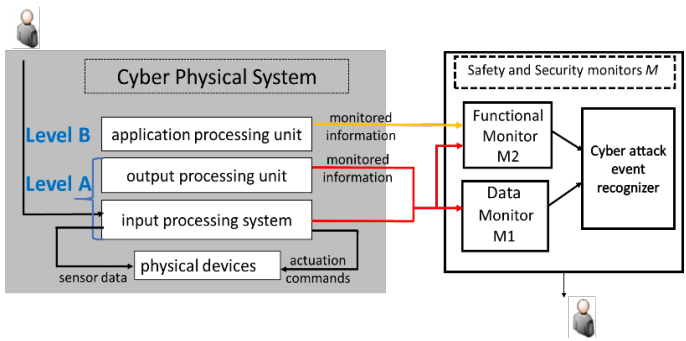

Fig. 2. Cyber Physical Systems with Multi-level Monitors.

## D. Rationale for multilevel monitoring

While the previous subsection extended monitoring from one monitor to a multiple monitor, this sub-section explains why we are locating monitors at different physical levels. Consider the simplified CPS in Fig. 2 that has monitors at two layers or levels: M1 monitors the data measured or sensor value and M2 monitors the functioning of a controller. An attack on level A (Fig. 1) is detected by monitor M1 at that level but is typically undetected by a monitor M2 (Fig. 2) at the level B. Similarly, an attack on the level B is detected by M2 but is typically undetected by M1 in the level A. Therefore, unless there are monitors at both these levels (irrespective of whether one can have single or multiple monitors at each level), a potentially dangerous attack may go undetected. Thus, having multiple such local monitors at each critical level helps provide more comprehensive coverage of the attack surfaces.

## E. Threat Model for Autonomous Vehicles

Autonomous vehicles which rely heavily on a variety of sensors and network connectivity for their self-driving intellect, poses several attack surfaces that can be exploited by attackers to compromise their safety and the data privacy of passengers. GPS, Lidar and Camera sensors, Bluetooth connectivity, Infotainment systems, On Board Diagnostics (OBD) port and the vulnerable ECUs themselves are potential attack surfaces in Autonomous vehicles [11]. Autonomous vehicles rely a lot on Lidars and Camera sensors for sensing its environment and making driving decisions and hence these attack surfaces are a major threat to Autonomous Vehicle safety. Lidar spoofing which is possible with an attacker's precisely controlled light source can lead to misguiding the vehicle about obstacles around it thereby resulting in unexpected vehicle behavior including emergency braking and frozen vehicle conditions [12]. Spoofing the camera sensors could cause vehicles to misinterpret traffic signs and speed limits resulting in dangerous driving conditions for passengers. Attacks on the GPS through GPS spoofing (which involves false signals of the same structure and frequency as of the authentic GPS signals being transmitted by an adversary's transmitter equipment) can have several impacts including giving incorrect information to the camera and causing several problems including diverting the vehicle from its intended route.

Man-in-the-middle attacks can exploit the On-board diagnostics (OBD) port in the cars to make the ECUs operate incorrectly or erase the stored information in ECUs by sending special commands through CAN bus and even reprogram the ECUs with malicious code. In this way, attackers can accomplish Denial of Service attack by disabling the ECU's participation in the CAN bus communication through unauthorized usage of commands on OBD II port. Bluetooth interface along with an insecure infotainment system in a vehicle is another potential attack surface that allows attackers to access private information in devices connected or paired with the vehicle Bluetooth. Hackers try to exploit software vulnerabilities such as Buffer Overflow, missing stack defenses, improper error handling etc. in Electronic Control units to feed in malicious inputs and affect their operation or the CAN bus transmissions through indirect physical (USB, OBD) or remote access (WIFI, Bluetooth) [10] [13].

The Data monitors at the inputs and outputs help detect any incorrect data, signal delays and missed information due to signal spoofing or component failure. The Functional monitors at the ECUs would detect any attack such would cause change in system functionality. We will study our monitoring approach for a representative Autonomous Emergency Braking system where the plant and sensors are in MathWorks Simulink [14], while the controller is implemented on a STMicroelectronics CORTEX M4 microcontroller [15]. Runtime monitors are implemented on a FPGA to verify the working of the AEB system.

IV. REPRESENTATIVE CYBER PHYSICAL SYSTEM : AUTONOMOUS EMERGENCY BRAKING (AEB)

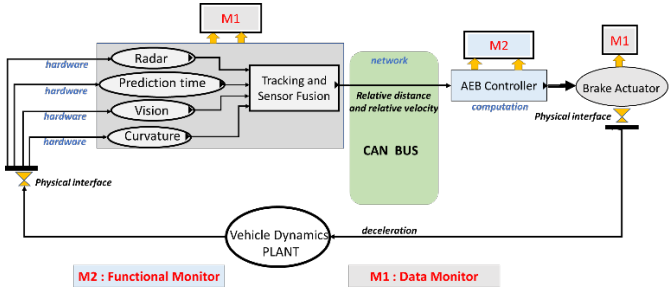

Fig. 3. Schematic of an Autonomous Emergency Braking (AEB) system showing the AEB controller, plant, sensors. Monitors that can be placed at various levels are also indicated.

Our representative system, which we monitor at multiple levels, is a simplified Autonomous Emergency Braking (AEB) system taken from the MathWorks Simulink examples library [14]. Fig. 3 shows an abstract view of the AEB system. Here the output of the AEB controller determines the braking state that decelerates the ego car (Ego car is the car with autonomous features). The dynamics of the car under this braking condition is modeled by the Vehicle Dynamics PLANT module whose output along with the scenario under consideration (explained later) determines the inputs to the radar and vision sensors. The output of these sensors are fused to estimate the relative distance and relatively between the ego car and Most Important Object (MIO). Note the MIO is not always the lead car. For example, if a pedestrian comes in front of the ego car, this would be the MIO.

Based on these inputs (such as relative distance and velocity to the MIO), the AEB controller estimates the braking state as summarized by Fig. 4(a). When the ego car is still at a safe distance but getting closer than a threshold for safe operation, an alert is issued. It the driver does not brake or the braking is insufficient, then at a certain critical relative distance, the AEB engages the stage I partial braking. If this does not suffice, a closer relative distance stage II partial braking is applied and then full braking is engaged. This decelerates the car to avoid collision. Avoiding near collision or a collision with a pedestrian is characterized by having a minimum headway distance when the velocity of the ego car reaches zero as shown in Fig. 4 (a).

The simulation scenario we consider is as following (from the Simulink library in [14]): The model plant (ego car) follows a lead car with a vehicle(s) on the side lane as shown in Fig. 4 (b). The vehicle(s) covers a pedestrian from the ego car's sensors till he/she crosses in to the lane of the ego car. The sensors: both camera and radar locate the pedestrian and the sensor information is fused to calculate the time needed to stop the car. When this is below a certain threshold, the driver is alerted and if sufficient braking action is not taken, as the object draws closer, the AEB initiates a first, then second stage partial braking, followed by full breaking. Thus, prevents collision with the target. In this paper, we perform monitoring at two levels to ensure the AEB system works correctly. First, we check the integrity of the sensor input using Data Monitors (DM) which ensure that the sensor inputs are within the safe operating bounds. Second, we verify the AEB controller functionality by Functional Monitors (FM) and verify that based on the inputs received by the AEB, the overall functionality satisfies all the properties of the AEB.

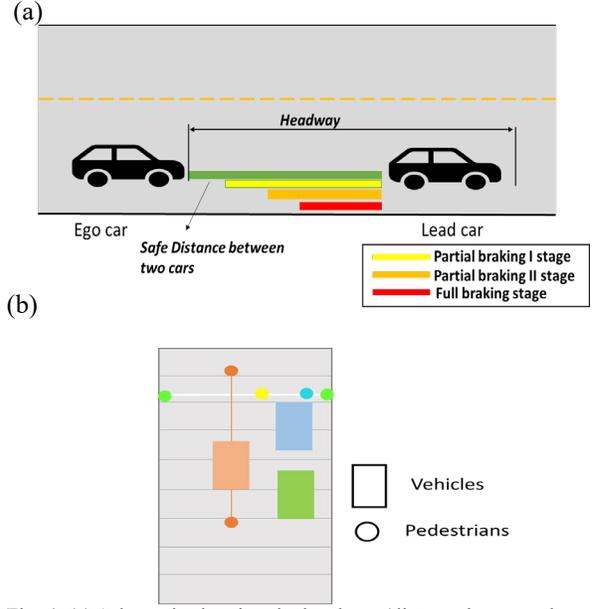

Fig. 4. (a) Schematic showing the headway (distance between the ego car and lead car) and various stages of autonomous breaking initiated. (b) Depiction of a scenario in the Simulink simulation (an example in their library that we used) where there are two vehicles ahead of the ego car but on the next lane to the right. They conceal a pedestrian who is crossing from right to left until he/she is in on the same lane as the ego car. Both figures are based on illustrations in Simulink that have been adapted and modified here.

Thus, our multilevel monitoring of the CPS spans: (i) an AEB (controller) implemented on Cortex M microprocessor; and (ii) the plant (car dynamics), sensor and sensor fusion modeled in MATLAB Simulink.

V. IMPLEMENTATION

We implement the runtime monitors for the AEB system initially at the model level in MathWorks Simulink where the monitors are used in FPGA in loop (FIL) mode [16]. Streams of information are input to the monitors from the Simulink model. The runtime monitors are generated using a stream-based runtime verification tool called TeSSLa [2]. After evaluating the monitors at the model level, we implement the AEB controller on an ARM Cortex M4 processor and verify the safety and security properties using TeSSLa monitors.

*A. Generation of Monitors*

The TeSSLa tool is ideally suited for specifying and verification of properties of CPS. This is because TeSSLa is a real-time specification language and operates on independent streams of data that are time-stamped. For example, consider three streams of data: p(t), q(t) and r(t). Since these are values of the data at different discrete points in time, we denote a stream as p (1), p (2), p (3) … p(n). Using TeSSLa, we can easily implement conditions such as the following:

If p(n) > q(n), then r(n+2) =5.

This simply means that if at any instant of time the data (value) of stream "p" is greater than that of stream "q", the value of the data in steam "r" after two instants should be "5". Thus, it is well suited for real-time online monitoring of relationship between various streams of data.

Such stream runtime verification (SRV) can check logical properties and compute temporal metrics and statistics from the trace [17]. There are many library functions in TeSSLa listed in [18] that allow such logical properties to be specified and synthesized on an FPGA to implement different monitoring conditions. An example of a TeSSLa specification for a requirement "increase or decrease in position $x$ over a time period of one second should be less than or equal to 5 m" is as shown in Fig. 5. This condition can be stated succinctly as:

$$|x(n) - x(n-1)| <= 5$$

We see that as $x$ goes from 1 m to 5 m between $t=0$ and 1 s and again from 5m to 10 m between $t=1$ and 2 s, this condition is met. Therefore, there is no attack (attack=false). However, as $x$ goes from 15 m and 100 m between $t=3$ and 4 s and again from 100 m to 20 m between $t=4$ and 5 s, this condition is not met. Therefore, there is an attack detected (attack=true) for two consecutive seconds. A snippet of the TeSSLa specification for this property is in Fig. 6.

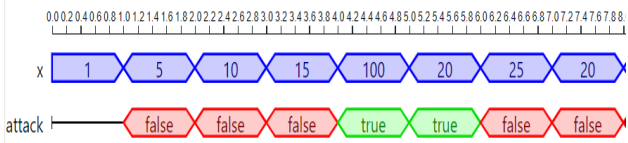

Fig. 5. Stream of data indicating position "x" in blue with time stamp (top) and monitor indicating whether an attack has occurred (bottom).

```
in x: Events[Int]
def attack:= x- prev(x) > 5 || x- prev(x) < -5
out x
out attack
```

Fig. 6. Snippet of the TeSSLa specification.

Verilog code synthesizable on FPGA can be generated from the TeSSLa specification and used as runtime monitors.

### B. Implementation and evaluation of Monitors

In this section, we first discuss the implementation of monitors on the FPGA that perform runtime monitoring of the AEB system running on Simulink. In the second part, we implement the AEB controller on an ARM Cortex M4 processor and directly monitor it's functioning with a monitor on FPGA, while the plant and sensor (continue to be simulated on Simulink as we do not have a physical car with sensors) and data integrity is also monitored by the FPGA.

#### 1) Hardware setup for AEB at Model Level

Both the plant with sensors and the AEB controller are implemented at the model level. Input and output streams from different sub-systems of this model are fed in real time to FPGA in the loop (FIL) as shown in the Fig. 7. These information streams are monitored on the XILINX Zynq-7000 XC7Z010-1CLG400C FPGA by implementing different monitors using TeSSLa.

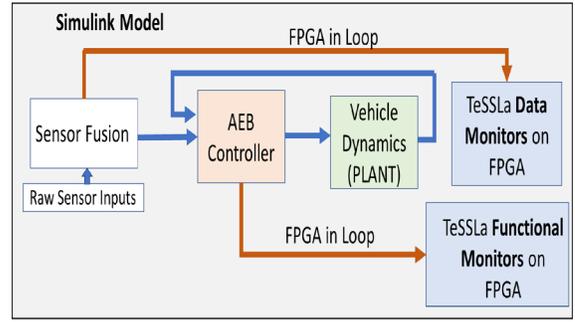

Fig. 7. Schematic showing the AEB controller, plant and sensors simulated on Simulink Model and Data Integrity of Sensors and Functionality of AEB Controller monitored by FPGA in the loop (FIL).

To keep the implementation simple, we chose one monitor at each level: Data and Functional. Attacks that can happen at each of these levels and the specific property monitored are described. Finally, we demonstrate why it is necessary to have separate monitors at both these physical levels.

*a) Data Monitor:*

The AEB system uses a combination of vision and radar sensors to estimate dynamic quantities such as ego vehicle speed, relative speed and relative distance between the ego vehicle and MIO. These estimated quantities can be corrupted, the sensors can be attacked or spoofed. For example, [19] showed that relatively inexpensive LIDAR spoofing devices could trick it into "sensing" a non-existent obstacle 20 to 350 m from the LIDAR unit.

In this work, we do not get into the details of whether such an attack is specifically made on the physical sensor or at the sensor fusion or corruption of data elsewhere. As far as the attack on this AEB system is concerned, the result is an erroneous change in relative distance or velocity between the ego car and the MIO.

As an example, we change the relative velocity by performing two attacks, one at t=3.1 seconds and the other at 4.4 seconds as shown in Fig. 8. The AEB controller is able to compensate for this attack, we therefore do not see an appreciable change in the headway, and eventually a collision is avoided. However, even this attack (with relatively minor effect on the overall AEB system performance) is detected by our monitoring condition: Magnitude of change in velocity by more than 10 m/s in 1 second is due to an attack. Since, the sampling is performed every 0.1 second this translates to

$$|u(n+1) - u(n)| < 1 \text{ m/s}.$$

In other words, when continuously monitoring the stream of information for the estimate of relative velocity, any jump in velocity (increase or decrease) more than 1 m/s between two successive data points in this stream is indicative of an attack on the system. Fig 8. (d) shows that the Data monitor detects both the attacks on the relative velocity. However, neither of the attacks are detected by the Functional monitor (described next) as an attack on the relative velocity does not change the functionality of the AEB controller and is therefore, not detected by the Functional monitor.

*b) Functional Monitor:*

The AEB controller receives various inputs from the sensor fusion (e.g. relative distance and velocity) and uses it to calculate quantities such as the Time To Collision (TTC), partial breaking-1-time, partial breaking-2, etc. Based on the relationship between these input quantities, the AEB controllers calculates the level of braking (stage 1, 2, 3), if any, needed to prevent a collision. There are several attacks that can occur on the functioning of the AEB controller, but like in the sensor scenario, we are less concerned with the exact modality of the attack. Ultimately, the consequence of such attacks is that the AEB controller fails to function in its normal way. For example, the AEB controller would fail to apply the braking condition-2 (more deceleration) or complete braking, when the partial braking condition-1 (less deceleration) does not suffice to prevent a collision. This is shown in Fig 9. When the condition: TTC < 0 & |TTC|< PB2 stopping time is met, the controller should be applying partial braking-2. However, even though this condition is met at t=2.7 seconds and the condition is identified (Fig 9. d) the brake state fails to go from "1" to "2" (Fig 9. c). This leads to the headway eventually reaching zero, i.e. a collision occurs. Again, this attack is successfully detected by the monitoring condition as see in 9 (f). The monitoring condition is

TTC < 0 & | TTC |< PB2 => AEB Braking status >= 2.

Again, since this is a failure in the functional relationship between the input and output of the AEB controller and not a data attack (sudden change in relative velocity for example), it is not detected by the Data Monitor.

*c) Summary: Data and Functional Monitors*

We have shown that an attack on the sensor and actuator data can only be detected by a Data Monitor and similarly an attack on the functioning of the AEB controller is only detected by the Functional monitor. In this model, the AEB controller braking output is directly sent to the plant and the sensor outputs are directly sent to the AEB controller. In real cars, this information is transmitted through the CAN Bus and therefore a monitor is also required at the communication level to identify delays in transmitting information, etc.

*(2) Hardware Setup for AEB on ARM Cortex M4 processor*

We implement the AEB controller on a STMicroelectronics ARM Cortex M4 processor and directly monitor it's functioning using on monitors on the FPGA, while the plant and sensor (continue to be simulated on Simulink as we do not have a physical car with sensors). Data integrity is monitored by the FPGA as shown in Fig. 10 in Simulink FIL mode. Any internal data or system states that we may need for monitoring was obtained by instrumenting the source code. ARM Coresight

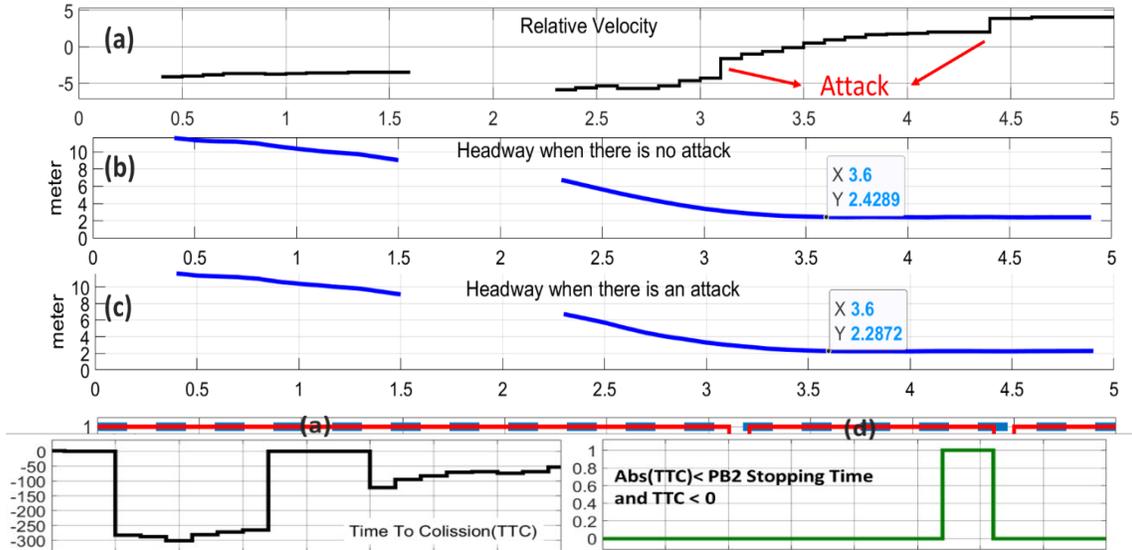

Fig. 8. Attack on relative velocity between ego car and MIO (b) Headway without attack and (c) with above attack (d) Both attacks on relative velocity detected by the Data monitor but are not detected by the Functional monitor.

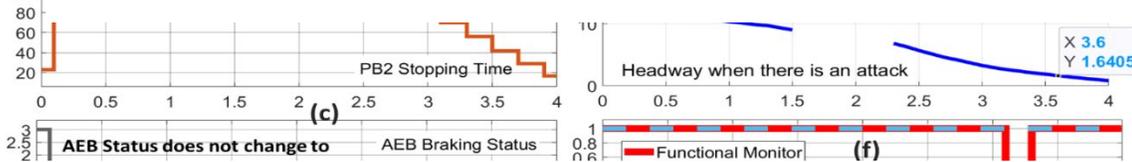

Fig. 9. (a) Time to collision (TTC) between ego car and MIO (b) Partial Breaking-2 (PB2) stopping time (c) AEB controller output that determines braking status (d) Checking if a relationship between TTC and PB2 holds (e) Headway approaches zero implying collision (f) Both attacks on relative velocity detected by the Data monitor but are not detected by the Functional monitor.

Architecture has a debug module called Instrumentation Trace

*d) Example of property monitored on the AEB controller*

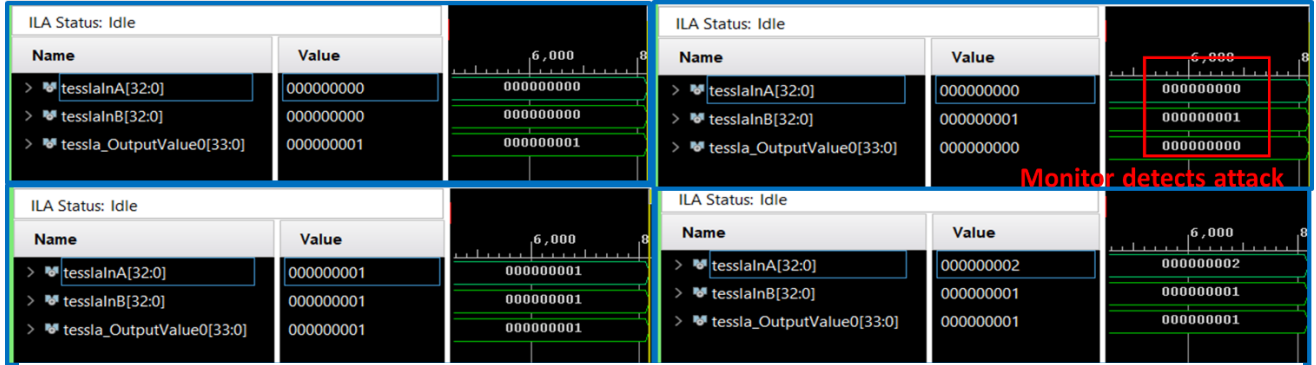

Fig. 11. Property based on data derived and decoded from ARM Core site ITM Trace. This checks for the condition that when AEB Status is 1, 2, 3 (i.e. not equal to "zero", the warning FCW Active must be "1". The top right sub-figure has AEB Status=1 but the FCW Active =0, indicating a violation of the property (case highlight in red). NOTE: I each sub-plot: tesslaInA corresponds to the input stream vale FW Active and tesslaInB corresponds to the input stream vale AEB Status, while tessla_OutputValue0 corresponds to the monitor. tessla_OutputValue0 (i.e. Monitor value =0 implies an attack has occurred).

Macrocell (ITM), which can be used to get actual values of variables that can be used to access system behavior with minimal intrusion to its working. ITM has 32 stimulus channels through which data to be monitored can be written. ITM can also perform Printf like debugging by writing data to the stimulus ports [20] [21].

The data obtained from the ITM module can be read either through a 4 pin Trace Port Interface Unit (TPIU) or through a Serial Wire Output (SWO) pin of a 2 pin Serial Wire Debug (SWD) port in the ARM processor [22]. In this paper, we use the SWD interface to read data traces due to ease of availability of decoders for this interface. But, one can always use the TPIU interface or the Aurora Gigabit Trace to get better performance. The Aurora trace unit can be used to get data of higher bandwidth more securely as it has Cyclic Redundancy Checks (CRCs) incorporated in them [23].

Fig. 10 shows the test bed that was used to monitor the traces on ARM processor. The STM32-MAT feature from MathWorks Simulink [24, p. 32] enables us to extract trace data for monitoring while the processor is running in Processor in Loop (PIL) mode in Simulink. In this setup, the processor communicates with Simulink through UART ports. The UART ports are used to receive sensor inputs and send controller output to the plant in Simulink.

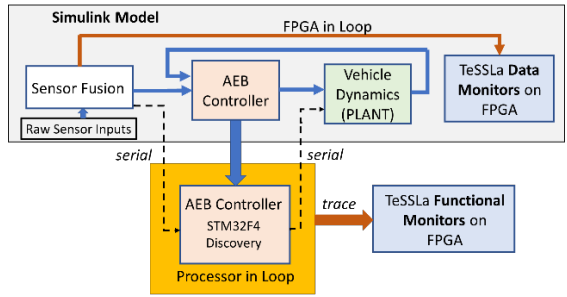

Fig. 10. Schematic showing the AEB controller implemented on an ARM Cortex M4 processor monitored by the FPGA, plant and sensors simulated on the Simulink model and monitored by the FPGA.

*using traces obtained from ARM processor*

The data traces obtained from ARM ITM module were decoded by the FPGA monitor and used for checking the properties. The results were viewed in the Integrated Logic Analyzer debug cores on the FPGA [25].

Consider a new monitoring condition that we implement here. It is based on the understanding that when the AEB controller is in any stage of partial or full braking (AEB status =1, 2, 3) the Forward Collision Warning (FCW Active) should be "1". We input FCW Active (top row in Table 1) and AEB Status (bottom row in Table 1) as shown. The monitoring condition is: [AEB Status = (1, 2 or 3)] => [FW Active =1]
Based on this condition the expected monitor outputs are shown in the bottom row of Table 1. FCW Active =0 when AEB Status =1 is erroneous and indicative of an attack that should result in Monitor =0.

Table 1. Input streams from Cortex M processor to FPGA and output stream from the FPGA monitor.

| FCW Activate | 0 | 0 | 1 | 1 |
|---|---|---|---|---|
| AEB Status | 0 | 1 | 1 | 2 |
| **Monitor** | 1 | 0 | 1 | 1 |

The Property based on data derived and decoded from ARM Core site ITM Trace in Fig. 11 shows that the TeSSLa monitor results viewed in Xilinx Integrated Logic Analyzer.

### C. Resources for FPGA implementation of Monitors

Resources utilized by a TeSSLa monitor is summarized in Table 2. This could vary depending on the specification. For our monitoring conditions, the slice registers account for less than 1% and LUTs account for less than 2% of the Xilinx Zynq-7000 XC7Z010-1CLG400C FPGA. Resource utilization with other blocks (ARM data trace decoders, not shown in the table) require negligible resources.

Table 2. Resource Utilization for one TeSSLa property

| Resource for TeSSLa Specification for 1 property | Used (% utilization expressed for Xilinx Zynq-7000 XC7Z010-1CLG400C) |
|---|---|
| Slice Registers | 316 out of 35,200 ( 0.9%) |
| LUTs | 319 out of 17,600 (1.81%) |
| Block RAMs | 0 of 60 ( 0 %) |
| DSP48 (multipliers) | 0 of 80 (0%) |

If we had 100 such ECUs in a car [26] of which 10% are safety critical, even monitoring all of these 10 safety critical ECUs will take less than 30% of the FPGA resources proving this concept is scalable. To prevent single point failures assuming three redundant monitoring conditions per ECU (instead of just one) the FPGA resource utilization would be less than 90%. A high-end FPGA [27] which has more resources than the Zynq-7000 XC7Z010-1CLG400C FPGA, would use only a small fraction of its resources. Such an FPGA could therefore manage more stringent, complex and resource intensive properties easily. This shows that the proposed multi-level monitoring concept is scalable.

## VI. CONCLUSION AND FUTURE WORK

We proposed a multilevel runtime monitoring approach where monitors are placed across different levels of a CPS to ensure timely detection and isolation of attacks. Such an approach was tested in MathWorks Simulink and implemented on FPGA hardware using stream-based runtime monitoring framework called TeSSLa. We use the ARM coresight debug and trace capability to extract data traces for monitoring. Finally, we show that this approach is scalable to a large number of ECUs.

In future, we would like expand monitoring to include execution monitoring to our multilevel monitoring framework using ARM Coresight Execution Trace Macrocell (ETM) and Program Trace Macrocell (PTM) modules to monitor control flow attacks. Further, we plan to verify the completeness of monitors using property based fault injection.

ACKNOWLEDGMENT

We would like to thank the TeSSLa runtime verification team, University Institute for Software Engineering and Programming Languages, University of Lubeck, Germany for assistance with the TeSSLa tool. (https://www.tessla.io/ )